\documentclass{INTERSPEECH2023}
\usepackage{multirow}
\interspeechcameraready 
\title{VoiceX: A Text-To-Speech Framework for Custom Voices}

\name{Silvan Mertes, Daksitha Withanage Don, Otto Grothe, Johanna Kuch, Ruben Schlagowski, Elisabeth André}

\address{
  Chair for Human-Centerd Artificial Intelligence, Augsburg University, Germany}
\email{silvan.mertes@uni-a.de}

\begin{document}

\maketitle

\begin{abstract}
Modern TTS systems are capable of creating highly realistic and natural-sounding speech. 
Despite these developments, the process of customizing TTS voices remains a complex task, mostly requiring the expertise of specialists within the field.
One reason for this is the utilization of deep learning models, which are characterized by their expansive, non-interpretable parameter spaces, restricting the feasibility of manual customization.
In this paper, we present a novel human-in-the-loop paradigm based on an evolutionary algorithm for directly interacting with the parameter space of a neural TTS model.
We integrated our approach into a user-friendly graphical user interface that allows users to efficiently create original voices. 
Those voices can then be used with the backbone TTS model, for which we provide a Python API.
Further, we present the results of a user study exploring the capabilities of VoiceX. We show that VoiceX is an appropriate tool for creating individual, custom voices.
\end{abstract}


\section{Introduction}
Text-to-Speech (TTS) systems are increasingly becoming an integral part of our daily lives. 
They seamlessly integrate into a variety of fields, like virtual agents, social media platforms, video games, and accessibility applications such as screen readers or navigation systems \cite{abdulrahman2022natural,mertes2021potential, zhang2023does,henderson2022screen,sangale2022literature}.
The speech quality of those TTS systems is steadily increasing, and the corresponding models are modified and extended in a variety of ways to equip them with even more capabilities, such as expressing voice-level emotions \cite{triantafyllopoulos2023overview} or even coping with multiple languages \cite{shang2021incorporating}.
One of those abilities that got heavily improved over the last years is the \emph{Multi-Speaker capability}. 
There, TTS models are not restricted to the synthesizing speech of one single voice. 
Instead, they are conditioned on multiple speakers' voices, so that they can produce a variety of different voice timbres \cite{casanova2022yourtts,chen2020multispeech,casanova2021sc,cooper2020zero}.

However, although those approaches possess the ability to generate a wide array of voices for different tasks, enabling users to fine-tune these voices to meet specific requirements presents a notable challenge.
Typically, a subset of voices, selected in advance by software engineers or sound designers, is made available to end-users.
If a user really wants to create a \emph{custom} voice that totally fits a certain use-case or need, there is commonly no possibility.
Even if the user has proficient technical knowledge, there is still another problem: in most state-of-the-art neural TTS models, the voice timbre is encoded in intermediate, latent representations inside the TTS model and gets implicitly learned during training.
As such, it is mostly not interpretable and therefore difficult to adapt.

\begin{figure}
\centering
  \includegraphics[width=1\linewidth]{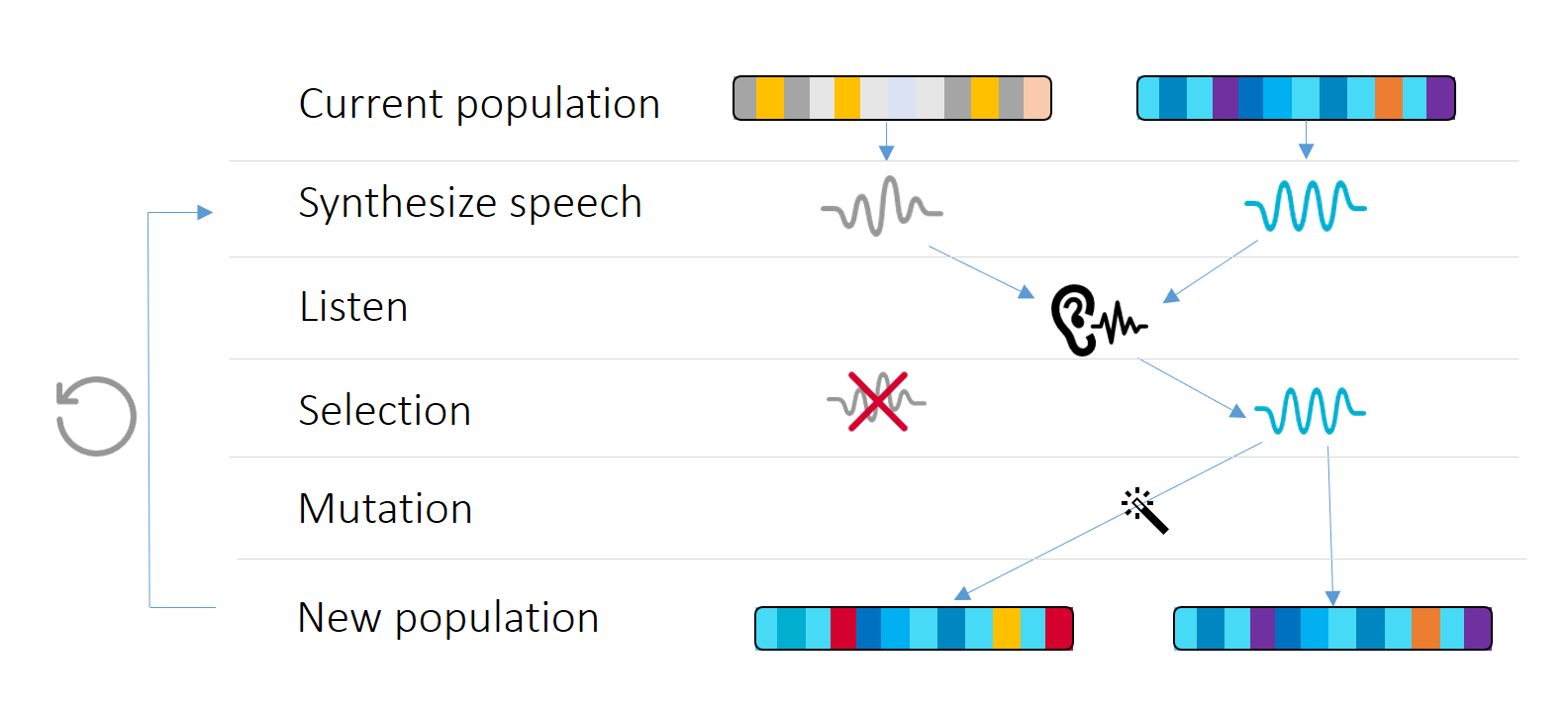}
  \caption{Schematic of the human-in-the-loop ES.}
  \label{fig:ea_scheme}
  %\vspace{-.5cm}
\end{figure}

In this paper, we present a human-in-the-loop paradigm to easily search through the latent speaker embedding space of the state-of-the-art TTS system \emph{VITS} \cite{kim2021conditional}.
Therefore, we introduce \emph{VoiceX}, a tool which iteratively presents users with pairs of distinct voices. 
Starting with an initial comparison, users are simply required to select the voice that ``fits better'' to their preference. 
These selections act as a feedback mechanism, forming the fitness function for an underlying Evolution Strategy (ES) algorithm (see Figure \ref{fig:ea_scheme}). 
Through this iterative process, VoiceX effectively navigates the complex space of voice timbres, refining the selections to converge towards a voice that aligns closely with the user's specific requirements.
%Therefore, we implemented a system, which we call \emph{VoiceX}, where the user repeatedly gets presented with two different voices, and only has to decide which one ``fits better''. 
%Those decisions serve as a fitness function for an underlying Evolution Strategy (ES) algorithm. 
%That ES searches through the latent speaker embedding space and has the goal to converge towards a speaker embedding that results in a TTS voice that perfectly fits the user's individual preference.
We integrated VoiceX into a human-in-the-loop web application to make voice customization easily accessible.
The system is designed to be easy to use, keeping cognitive load at a minimum, and as such being suitable to a broad spectrum of users.
We evaluated our system in a user study (N=65). 
There, we tested if users are capable of creating voices of certain characteristics. 
Specifically, as a proof-of-concept, we assessed whether VoiceX can be utilized to build voices that exhibit distinct personality traits. 
As personality is a quite complex concept, we argue that it therefore serves as an appropriate characteristic to explore the capabilities of the system.

The contributions of this work are as follows:
%\vspace{-.2cm}
\begin{itemize}
    \item We present a novel human-in-the-loop paradigm for customizing TTS voices.
    \item We publicly provide a web application that can be used to create new TTS voices.
    \item We present a study examining the voice customization capabilities of the system.
    \item We provide an open-source Python API for easy usage of the created voices in all sorts of applications.
\end{itemize}

\section{Related Work}

\subsection{TTS Voice Customization}
To customize voices in TTS systems, the underlying models must support such modifications.
%Customizing a voice for TTS systems inevitably implies that the corresponding TTS model allows to do so.
%To this end, various approaches exist to integrate customization mechanisms into the model.
Wang et al. presented \emph{Tacotron GST} \cite{wang2018style}, a modification of Tacotron 2 \cite{shen2018natural}. There, \emph{Global Style Tokens} are introduced, facilitating the direct manipulation of certain implicitly learned attributes, such as speech speed and speaking style.
Similarly, Valle et al. introduced Mellotron \cite{valle2020mellotron}, which was conditioned on rhythm and continuous pitch contours, allowing for the direct modification of those features after training. 
However, the number of features that can be adjusted in both those architectures is quite limited.
Although in theory, Tacotron GST gives the possibility to modify the style token representations on a fine granular level, those representations then do not hold interpretable semantics -- making them difficult to manually adjust.
The same goes for the \emph{VITS} model \cite{kim2021conditional}, which makes use of speaker embeddings. 
As such, by altering these embeddings, different voices are synthesized.
However, since speaker embeddings itself may not be inherently interpretable, manual adjustments remain a challenge.
Overall, these approaches share a common limitation: customization options are either confined to a limited number of features, or they remain inaccessible due to the non-interpretable nature of latent representations.
One tool that wraps various TTS systems in a way that different features can be steered by end users is \emph{So-to-Speak} \cite{szekely2023so}. 
However, their approach is more directed towards model capability \emph{analysis} and audiovisual \emph{exploration}, while VoiceX breaks down the TTS system's adjustment possibilities to binary decision tasks, focusing on simple and efficient \emph{customization}.
%have in common that the customization capabilities are either restricted to few features, or they are not accessible due to latent representations not being interpretable.

\subsection{Collaborative Voice Creation}
Contrary to the approaches above, some approaches have been presented in the past where customization of voice was performed not on an individual basis, but collaboratively, i.e., a group of people adapting a voice together. 
For example, Harrison et al. presented the concept of \emph{Gibbs Sampling with People} (GSP), in which experiment participants were able to manipulate specific, interpretable voice features, such as pitch range and F0 perturbation, allowing the voice to convey emotional prosody \cite{harrison2020gibbs}. 
That concept was followed up by van Rijn et al. in multiple studies where they extended the GSP paradigm to work with latent spaces of neural TTS systems to create emotional voice prototypes \cite{van2021exploring}, voices for human faces \cite{van2022voiceme}, and voices for robots \cite{van2024giving}.
In contrast to VoiceX, all those approaches focus on collaborative adaptation, leading to more generic prototypes of voices instead of \emph{individually} customizing a voice to a \emph{single} user's preferences. 

\subsection{Voice Cloning}

Voice cloning utilizes algorithms -- typically rooted in deep learning -- to replicate a particular individual's voice, aiming to mimic the speaker's vocal traits, intonations, and speech patterns. 
%To achieve a reasonable likeness to the original voice, this process usually requires a substantial amount of training data from the targeted individual to capture their unique vocal nuances.
Recent advances have brought a variety of approaches to voice cloning, mostly focusing on specific factors like real-time capabilities \cite{jemine2019real}, expressiveness \cite{neekhara2021expressive}, speech quality \cite{hu2023real}, or the ability to clone with little training data \cite{arik2018neural}.
%In contrast, crafting a customizable voice centers on producing a voice that isn't necessarily an imitation of an existing individual. 
%Essentially, voice cloning focuses on imitation, while customizable voice creation revolves around tailoring and innovating \emph{original} and/or \emph{new} voices.
In contrast, the goal of creating \emph{custom} voices (as we aim for with VoiceX) is not to replicate, but to innovate, allowing for the generation of \emph{original} or \emph{novel} voices tailored to specific requirements or preferences.

\section{The VoiceX Framework}
\subsection{Backbone TTS Model}

\begin{figure*}
\centering
  \includegraphics[width=1\textwidth]{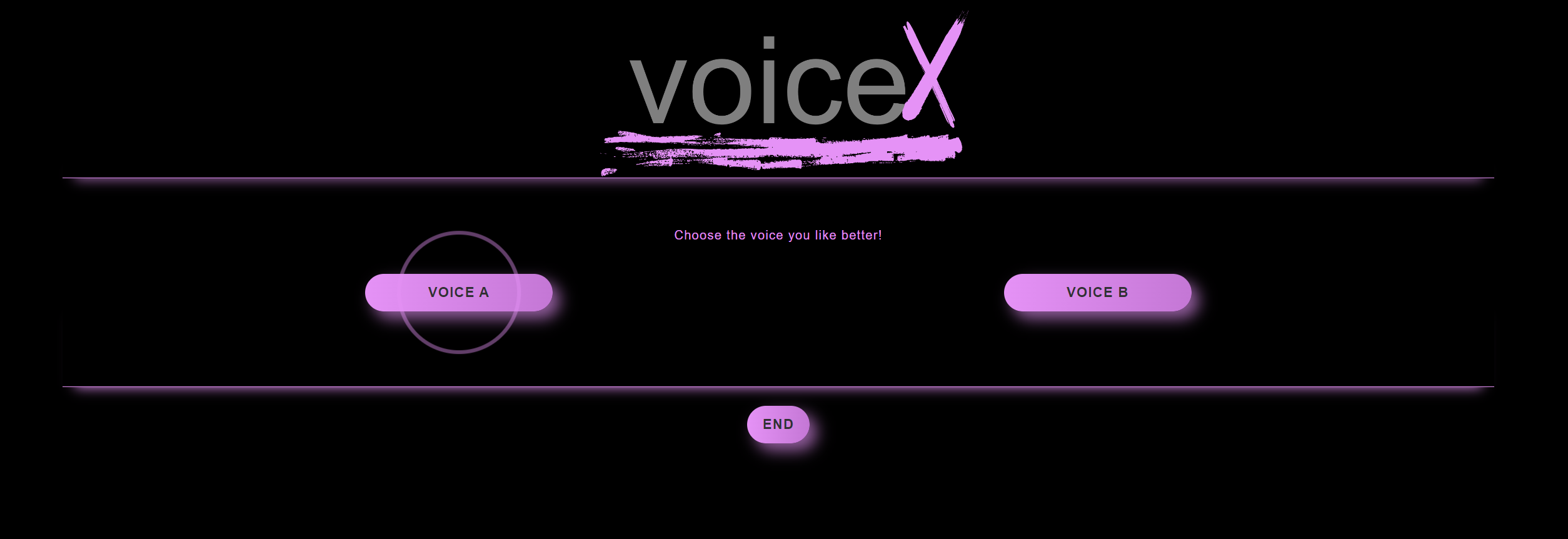}
  \caption{Screenshot of the VoiceX web application.}
  \label{fig:screenshot_gui}
  %\vspace{-.5cm}
\end{figure*}

We use \emph{VITS}  \cite{kim2021conditional}, a state-of-the-art neural TTS architecture, as backbone for the whole system. 
Specifically, we use a model\footnote{The pretrained model was taken from \url{https://github.com/jaywalnut310/vits}.} trained on the VCTK dataset \cite{yamagishi2019cstr}, as that model has shown its ability to cover a wide variety of voices in related experiments \cite{van2022voiceme,van2024giving}.

\subsection{Speaker Representation}
An integral part of the VITS model is the use of so-called \emph{speaker embeddings}. 
Those embeddings are a 256-dimensional representation of the speaker's voice, and its semantics were implicitly learned during the training of the model.
As such, feeding different speaker embeddings with the synthesis text produces varied voices. Therefore, by interpolating the speaker embedding space we can generate a broad variety of different voices.
To reduce the search space of the following human-in-the-loop algorithm (as described in the next section), we use a principal component analysis (PCA).
Specifically, we use the PCA model that has already been used by van Rijn et al. \cite{van2024giving}. Their model was fitted on the 110 speakers of the VCTK dataset. 
We perform our human-in-the-loop algorithm solely on the first 10 PCA components of that model, which we inversely feed into the PCA model to obtain the 256-dimensional representation that the VITS model demands.

\subsection{Voice Creation} \label{sec:voice_creation}
\textbf{Evolution Strategy (ES).}
The modification of the synthesized voice is achieved through the implementation of a $(1 + \lambda)$-ES \cite{beyer2002evolution}, where $\lambda$ represents the offspring size. 
As such, one whole population consists of $1 + \lambda$ individuals. 
One individual contains the first 10 PCA component values of one speaker embedding.
In common evolutionary strategies, the phenotypes (which in our case would be voices synthesized using the speaker embeddings obtained through the individual PCA component sets) of the whole population would be evaluated by a fitness function to determine the best individuals. 
For our human-in-the-loop approach, that fitness function gets replaced by human judgment, i.e., the user assesses the best individual by listening to the voices.
A similar approach was followed by Ritschel et al. \cite{ritschel2019personalized} -- although they did not tackle the use-case of TTS customization. Rather, they used a human-in-the-loop ES to steer the customization of non-verbal affective sounds for social robots.
We set $\lambda$ to $1$ -- which means that in total, one population consists of one parent individual and one offspring individual. 
As such, the user must only listen to two voices per iteration and judge which one s/he likes better.
After that decision has been made, the PCA set that produces the better results is selected as the parent for a new population.
The offspring of the new population is produced by mutating that parent. 
For mutation, we implemented an epsilon mutation strategy with $\epsilon = 0.2$, which means that in $80\%$ of the cases, the offspring is produced by an actual mutation of the parent, and in $20\%$ of the cases the offspring is just a random new individual. 
Doing so prevents the whole ES loop from getting stuck in a local minimum. The mutation operation itself adds Gaussian noise to the PCA components, leading to a slightly altered voice.
After the mutation took place, the new population (i.e., the best individual from the last iteration and newly formed individual) is transformed into speaker embeddings (by the inverse use of the PCA model), synthesized to speech (by the use of the VITS model), and the resulting phenotypes of the new population are again presented to the user.
The overall loop continues until the user is satisfied with a voice.
As related work \cite{van2022voiceme} has shown that the most important characteristic that steers voice customization decisions is the perceived gender of voices, we wanted to make sure that users can set that course right from the beginning of the interaction.
As such, for the initial population, we use two predefined speaker embeddings: one that represents a stereotypical male voice, and one that represents a stereotypical female voice. 
Figure \ref{fig:ea_scheme} depicts the overall scheme of our human-in-the-loop evolutionary algorithm.

\textbf{Web Interface.}
To emphasize the overall simplicity of our voice creation paradigm, we designed an easy-to-use web interface that can be used to perform the human-in-the-loop evolutionary strategy\footnote{URL blinded for review.}. 
A screenshot of the interface is shown in Figure \ref{fig:screenshot_gui}. 
As can be seen there, the interaction elements are two buttons that can be used to decide on the preferred voice of the current iteration. 
The voices are played back alternately, with visual animations indicating the voice that is played.
An additional button is used to end the process. 
After clicking that button, the user can end the voice creation loop and download a voice file -- that voice file can then be used for TTS purposes through our dedicated open-source Python API that we provide publicly as PyPi package\footnote{Link to the repository will be made publicly available on acceptance.}.

\section{Evaluation}
We conducted a user study to evaluate if users are able to build custom voices with our system. 

\subsection{Study Design}

\begin{table*}[]
\begin{tabular}{l|l|l|l|l|l|}
\cline{2-6}
                                                                                                  & \textbf{Mean} & \textbf{Median} & \textbf{Std} & \textbf{Min} & \textbf{Max} \\ \hline
\multicolumn{1}{|l|}{\textit{"How did creating a voice with the given personality work for you?}} & 4.57          & 5               & 0.56         & 3            & 5            \\ \hline
\multicolumn{1}{|l|}{\textit{"How would you assess the voice quality of your created voice?"}}    & 4.32          & 4               & 0.47         & 4            & 5            \\ \hline

\end{tabular}

\caption{Results for the subjective measures.}
%\vspace{-.5cm}
\end{table*}

In our study, we focused on creating voices with a certain \emph{personality}. 
Personality is a spectrum of different traits (often defined by the \emph{Big Five} model \cite{goldberg2013alternative}), and personality in voices might be perceived quite differently among people.
Therefore, we argue that the task of creating voices with certain personalities can serve as a good proxy for assessing if our system can be used to create custom, personalized voices.
As such, participants were tasked with creating a voice that either accentuated or subdued one specific trait of the Big Five dimensions \emph{Extraversion}, \emph{Agreeableness}, \emph{Conscientiousness}, \emph{Neuroticism} or \emph{Openness}.
This assignment yielded ten unique voice creation tasks, where each participant was randomly allocated to \emph{one} of those.
%As such, each participant had the task to build a voice that shows one specific personality trait. 
%Here, we used the Big Five model, which consists of the five traits \emph{Extraversion}, \emph{Agreeableness}, \emph{Conscientiousness}, \emph{Neuroticism} and \emph{Openness} \cite{goldberg2013alternative}. 
%Each participant had the task to create a voice that either maximized or minimized one of the traits, leading to ten different tasks. 
%Each participant was randomly assigned one of those tasks. 
For phrasing the tasks, we used the terminology of John et al., who introduced adjectives both for minimizing and maximizing specific traits, e.g., \emph{Neurotic} for maximized neuroticism opposed to \emph{emotionally stable} for minimized neuroticism \cite{john1999big}.
After that, participants had to fill out questionnaires regarding their created voice. 
The first question that we wanted to explore was: \emph{How well can users create voices that show certain characteristics?}
To answer that, we presented the users with the Big Five Inventory (BFI) questionnaire \cite{john1991big}, where they had to assess how much their created voice matches with the presented items. 
Specifically, users had to fill out only the subscale of the questionnaire that matches the personality trait of their given task. 
As the items of that questionnaire do not directly use the terminology that we used for the task descriptions, but rather paraphrase them by multiple correlating items, that questionnaire can be seen as a semi-objective measure -- participants did not directly rate how well they performed in their task, but rather assessed characteristics of the created voice.
As the objective evaluation of the voices does not necessarily have to correlate with the subjective self-assessment of the users, we additionally asked the participants directly to rate how creating a voice with the given personality worked for them (Likert-Scale: 1=Did not work at all, 5=Worked very well).
Lastly, we asked the participants to rate the quality of their created voice (Likert-Scale: 1=Bad, 5=Excellent), before giving the option to give free-form textual feedback regarding both the system and the created voice.
The study was conducted online through Amazon MTurk. 
We acquired 65 participants, out of which 41 were male and 24 were female. 
Subjects were between 22 and 64 years old (M=35, avg=33.66, SD=7.08). 

% \usepackage{multirow}
\begin{table}[]
\begin{tabular}{ll|ll|}
\cline{3-4}
                                                    &                      & \multicolumn{2}{l|}{\textbf{Score}}  \\ \cline{3-4} 
\textbf{BFI Subscale}                               & \textbf{Task}        & \multicolumn{1}{l|}{\textbf{Mean}} & \textbf{Std} \\ \hline
\multicolumn{1}{|c|}{\multirow{2}{*}{Extraversion}} & \textit{Extroverted} & \multicolumn{1}{l|}{3.77}          & 0.51            \\ \cline{2-4} 
\multicolumn{1}{|c|}{}                              & \textit{Introverted} & \multicolumn{1}{l|}{3.25}          & 0.125            \\ \hline
\multicolumn{1}{|c|}{\multirow{2}{*}{Agreeableness}} & \textit{Agreeable} & \multicolumn{1}{l|}{3.60}          & 0.67            \\ \cline{2-4} 
\multicolumn{1}{|c|}{}                              & \textit{Antagonistic} & \multicolumn{1}{l|}{3.40}          & 0.49            \\ \hline
\multicolumn{1}{|c|}{\multirow{2}{*}{Conscientiousness}} & \textit{Conscientious} & \multicolumn{1}{l|}{3.38}          & 0.49            \\ \cline{2-4} 
\multicolumn{1}{|c|}{}                              & \textit{Lacking of direction} & \multicolumn{1}{l|}{3.28}          & 0.47            \\ \hline
\multicolumn{1}{|c|}{\multirow{2}{*}{Neuroticism}} & \textit{Neurotic} & \multicolumn{1}{l|}{2.65}          & 0.29            \\ \cline{2-4} 
\multicolumn{1}{|c|}{}                              & \textit{Emotionally stable} & \multicolumn{1}{l|}{2.55}          & 0.64            \\ \hline
\multicolumn{1}{|c|}{\multirow{2}{*}{Openness}} & \textit{Open to experiences} & \multicolumn{1}{l|}{3.52}          & 0.21            \\ \cline{2-4} 
\multicolumn{1}{|c|}{}                              & \textit{Closed to experiences} & \multicolumn{1}{l|}{3.82}          & 0.25            \\ \hline
\end{tabular}
\caption{Results for the semi-objective measures.}
\end{table}

\subsection{Results}
Results for the subjective measures are shown in Table 1.
Here, the self-assessment of how creating a voice with the given personality worked for the participants averaged at $4.57$, while the mean opinion score regarding the voice quality of the created voice reached a mean of $4.32$.

The results for the semi-objective part of the study (i.e., the BFI subscale scores) are shown in Table 2. 
%There, each boxplot aggregates the results for one of the ten different tasks that were randomly assigned. 
%Each row contains one of the big five traits, while the first column represents the task for creating a voice that maximizes that trait, while the second column represents the task for minimizing the same trait.

\section{Discussion}

\begin{figure}
\centering
  \includegraphics[width=1\linewidth]{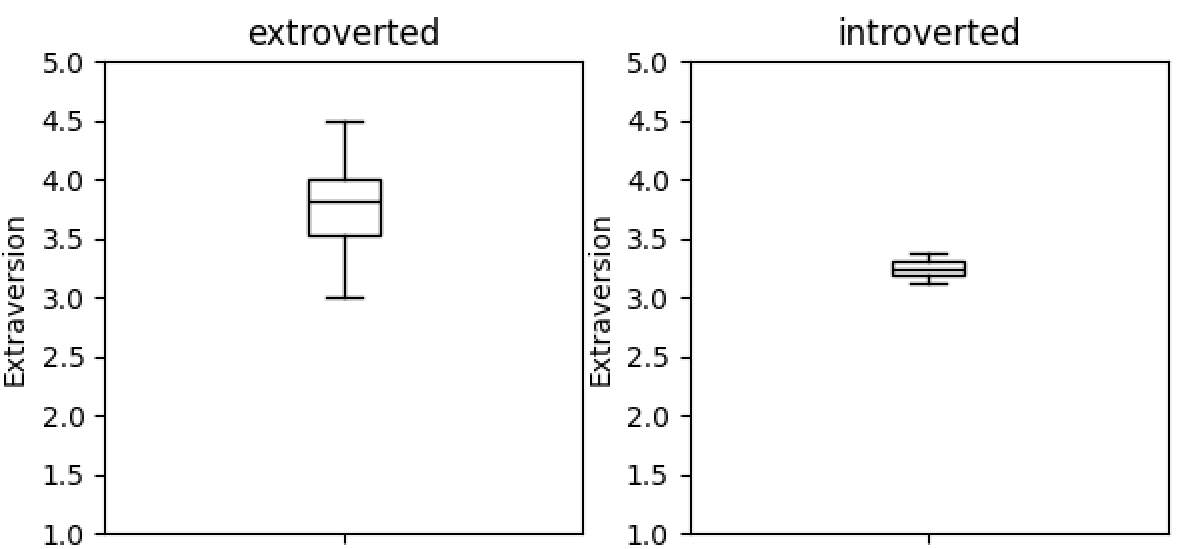}
  \caption{Big Five Inventory (BFI) extraversion subscale.}
  \label{fig:bigfive}
\end{figure}
The outcomes of our study highlight several pivotal insights into the VoiceX framework's effectiveness and the subjective nature of voice personality perception.

First and foremost, the voice quality mean opinion score (MOS) was $4.32$, which is very similar to the MOS that was reported by the authors of VITS (which was, depending on the used evaluation dataset, between $4.38$ and $4.43$). 
This underscores VoiceX's ability to maintain high speech quality, despite the interpolation through the speaker embedding space is performed without further constraints to maintaining a certain value distribution.
This finding is critical, as it demonstrates that our customization approach does not compromise the inherent quality of the synthesized voices.
%, voices can be created where the speech quality is not substantially worse than the original VITS system.

As can be seen in the results regarding the customization capabilities of VoiceX, there is a huge discrepancy in the semi-objective versus subjective measures. 
While in the former one (see Table 2), there is no tendency observable that participants overall were able to create a voice that matches the personality which their task demanded, the subjective measure tells the opposite: the explicit question of how creating a voice with the given personality worked for the participants was rated with $4.57$ out of $5$ on average.
This divergence between semi-objective and subjective assessments of the voices is majorly interesting and inevitably leads to the question of how this could happen.
One explanation might be the chosen personality questionnaire for the semi-objective assessments of the voices. 
The BFI was initially developed to self-assess personality. 
We have ``misappropriated'' it as a measure to assess the personality of a voice -- which might not have been the optimal choice. 
Another possible reason for the contrary results might be that participants simply do not have an imagination of how a voice with a specific personality sounds, as the voice itself might not directly correlate with certain personality traits. 
This is in line with findings from Scherer et al. \cite{scherer1978personality} who reported that in an experiment where humans labeled speech regarding certain personality traits, they could find significant correlations between speech and personality only for extraversion -- which is also the only personality trait where we could observe a notable and correct trend between the trait minimization and maximization tasks, as depicted in Figure \ref{fig:bigfive}.
As such, the interaction with the VoiceX system might have been rather exploratory, maybe even ignoring the overall target -- but still having the feeling of \emph{doing it right}. 

As the participants' subjective measure of how they performed was high, we claim that the system has huge potential for voice customization: After all, the own \emph{perception} of how the task worked out is arguably the most important metric when evaluating the efficacy of customizing a voice to one's \emph{own}, \emph{personal} needs.

\section{Conclusion}
In this work, we presented \emph{VoiceX}, a system that uses a \emph{VITS} model as the backbone TTS engine and allows users to create custom voices in an efficient, user-friendly way.
VoiceX is based on a human-in-the-loop evolutionary strategy, where the user's decision serves as the fitness function. 
By integrating our approach into a web application, we provide an accessible, easy-to-use tool that everyone can use to create personalized voices.
Additionally, we made a Python API publicly available, allowing for an integration of the created voices into all kinds of applications.
A user study showed that VoiceX can be used to create voices of high quality, as indicated by a MOS of $4.32$.

While semi-objective measures of the user study indicated that participants were not able to use the system to create voices that show a certain personality, users still had the impression that they were able to do so. 
We discussed that this deviation between semi-objective and subjective measures might stem from the users not having a detailed imagination of what a voice with a certain personality trait sounds like. 
However, the subjective measure of how creating a voice with a certain personality worked was extremely positive, indicating that participants could effectively apply the system to their own satisfaction. 

Overall, we argue that the perception of being able to customize a voice in the desired way is the key requirement for personalized, individual voice customization. Therefore, we can conclude that VoiceX shows great potential for a variety of stakeholders and use cases.

%\section{Acknowledgements}

\bibliographystyle{IEEEtran}
\bibliography{main}

\end{document}